\documentclass[onecolumn,sigconf]{aamas}  % do not change this line!

\usepackage{booktabs}
\usepackage{times}  %Required
\usepackage{helvet}  %Required
\usepackage{courier}  %Required
\usepackage{url}  %Required
\usepackage{graphicx}  %Required

\usepackage{amsmath}
\usepackage{algorithm} 
\usepackage[noend]{algpseudocode}
\usepackage{float}
\usepackage{soul}
\usepackage{amssymb}
\usepackage{tabu}
\usepackage{array}
\usepackage{multirow}
\usepackage{mathtools}
\usepackage{url}
\usepackage{color}
\usepackage{caption}
\usepackage{subcaption}
\usepackage{epstopdf}
\usepackage{hyperref}
\newcommand\given[1][]{\:#1\vert\:}
\algnewcommand\algorithmicforeach{\textbf{for each}}
\algdef{S}[FOR]{ForEach}[1]{\algorithmicforeach\ #1\ \algorithmicdo}
\algdef{S}[If]{ForEach}[1]{\algorithmicforeach\ #1\ \algorithmicdo}
\DeclareMathOperator*{\argmax}{arg\,max}

\setcopyright{ifaamas}  % do not change this line!
\acmDOI{doi}  % do not change this line!
\acmISBN{}  % do not change this line!
\acmConference[AAMAS'18]{Proc.\@ of the 17th International Conference on Autonomous Agents and Multiagent Systems (AAMAS 2018), M.~Dastani, G.~Sukthankar, E.~Andre, S.~Koenig (eds.)}{July 2018}{Stockholm, Sweden}  % do not change this line!
\acmYear{2018}  % do not change this line!
\copyrightyear{2018}  % do not change this line!

\acmPrice{}  % do not change this line!

\begin{document}
\title{Combining Prediction of Human Decisions with ISMCTS in Imperfect Information Games}  % put your title here!
%\titlenote{Produces the permission block, and copyright information}
\titlenote{This work was supported in part by the LAW-TRAIN project that has received funding from the European Union Horizon 2020 research and innovation program under grant agreement 653587 and by MAAFAT. In addition, a grant from the Ministry of Science \& Technology, Israel \& the Japan Scienceand Technology Agency (jst), Japan }
% AAMAS: as appropriate, uncomment one subtitle line; check the CFP
%\subtitle{Extended Abstract}
%\subtitle{Industrial Applications Track}
%\subtitle{Socially Interactive Agents Track}
%\subtitle{Blue Sky Ideas Track}
%\subtitle{Robotics Track}
%\subtitle{JAAMAS Track}
%\subtitle{Doctoral Mentoring Program}

%\subtitlenote{The full version of the author's guide is available as \texttt{acmart.pdf} document}

% AAMAS: submissions are anonymous for most tracks
% \author{Paper \#446}  % put your paper number here!
% Moshe Bitan\\
% Department of Computer Science\\
% Bar-Ilan University, Israel\\
% moshe.bitan78@gmail.com
% \And 
% Sarit Kraus\\
% Department of Computer Science\\
% Bar-Ilan University, Israel\\
% sarit@cs.biu.ac.il\\

\author{Moshe Bitan}
% \authornote{Dr.~Trovato insisted his name be first.}
% \orcid{1234-5678-9012}
\affiliation{%
 \institution{Department of Computer Science \\ Bar-Ilan University, Israel}
%  \streetaddress{P.O. Box 1212}
%  \city{Ramat-Gan} 
%  \state{Israel} 
%  \postcode{43017-6221}
}
\email{moshe.bitan78@gmail.com}

\author{Sarit Kraus}
% \authornote{Dr.~Trovato insisted his name be first.}
% \orcid{1234-5678-9012}
\affiliation{%
 \institution{Department of Computer Science \\ Bar-Ilan University, Israel}
%  \streetaddress{P.O. Box 1212}
%  \city{Ramat-Gan} 
%  \state{Israel} 
%  \postcode{43017-6221}
}
\email{sarit@cs.biu.ac.il}

\begin{abstract}

\textit{Monte Carlo Tree Search} (MCTS) has been extended to many imperfect information games.
However, due to the added complexity that uncertainty introduces, these adaptations have not reached the same level of practical success as their perfect information counterparts. In this paper we consider the development of agents that perform well against humans in imperfect information games with partially observable actions. We introduce the Semi-Determinized-MCTS (SDMCTS), a variant of the Information Set MCTS algorithm (ISMCTS). More specifically, SDMCTS generates a predictive model of the unobservable portion of the opponent's actions from historical behavioral data. Next, SDMCTS performs simulations on an instance of the game where the unobservable portion of the opponent's actions are determined. Thereby, it facilitates the use of the predictive model in order to decrease uncertainty. We present an implementation of the SDMCTS applied to the Cheat Game, a well-known card game, with partially observable (and often deceptive) actions.
Results from experiments with 120 subjects playing a head-to-head Cheat Game against our SDMCTS agents suggest that SDMCTS performs well against humans, and its performance improves as the predictive model's accuracy increases. 

\end{abstract}
  % put your semicolon-separated keywords here!

\maketitle

\pagestyle{empty}
\section{Introduction}

%Monte Carlo Tree Search (MCTS) is a simulation based search algorithm that is designed to handle huge search spaces by  exploring the most auspicious regions of the state space.
Monte Carlo Tree Search (MCTS) has had significant success in perfect information games such as Scrabble and Hex \cite{sheppard2002world,arneson2010monte} and most noticeably in computer Go \cite{gelly2012grand,gelly2006modification,silver2016mastering}. In recent years, MCTS was adapted by researchers \cite{browne2012survey} to solve imperfect information games.
%The game state in imperfect information games is not fully observable by the players. 
However, due to the added complexity that uncertainty introduces, these adaptations have not reached the same level of practical success as their perfect information counterparts. 
%This is partially due to the limited use of learning methods for reducing uncertainty. 
This paper focuses on developing agents that can play well against humans in imperfect information games by applying learning models to reduce uncertainty. More specifically, we focus on predicting the unobservable portion of the human opponent's actions and combining the prediction into our novel MCTS adaptation.

One of the first popular extensions of MCTS to imperfect information games is a MCTS variation where a perfect information search is performed on a determinized instance of the game \cite{ginsberg2001gib,sheppard2002world,sturtevant2008analysis,bjornsson2009cadiaplayer}.
That is, the search is performed on an instance of the game where hidden information is revealed to the players, thereby transforming the imperfect information game to a perfect information game. 
This approach allows the deployment of a predictive model. However, several problems have been reported when applying determinization which prevent the convergence to an optimal strategy such as \textit{Strategy Fusion} as discussed later in section "Determinization" \cite{ponsen2011computing}.

To address these problems, Information set MCTS, a new family of MCTS adaptations, were developed \cite{cowling2012information,heinrich2015smooth,lisy2015online}. ISMCTS and its variants perform MCTS simulations directly on information sets. An information set is a set of game states that are indistinguishable by the player at a given stage of the game. However, ISMCTS assumes that every game state in the information set has the same probability of being the game state which is not always the case. Thus, it prevents the use of a predictive model.% which reduce uncertainty.

This paper proposes a method for the development of agents playing against humans in imperfect information games and partially observable actions. We introduce the Semi-Determinized-MCTS (SDMCTS), a variant of the Information Set MCTS algorithm (ISMCTS) combined with a determinization technique to take advantage of the benefits of both approaches. SDMCTS is designed to reduce uncertainty by utilizing a predictive model of the unobservable portion of the opponent's actions. Similar to ISMCTS, SDMCTS performs simulations directly on information sets with one important distinction. The SDMCTS performs simulations on an instance of the game where the unobservable portion of the opponent's actions are determined. The opponent's actions are determined while keeping the remaining unobservable information hidden (e.g. opponent's cards). Thereby, it preserves the advantages of ISMCTS (i.e. reduce Strategy Fusion) while facilitating the use of the predictive model for decreasing uncertainty with regards to hidden actions.

We evaluate the SDMCTS in \textit{Cheat Game}, a well-known card game. The players' goal in Cheat Game (also known as Bullshit and I Doubt It) is to discard the initially dealt cards. At each turn, a player may discard several cards and declare their rank. However, the cards are placed  facedown and the player is allowed to lie about their rank. Thereby, the action is only partially observable by the opponents. For the development of the SDMCTS agents, a behavioural-based predictive model of human player actions in the Cheat Game was used. The predictive model was trained on data collected from 60 players playing human-vs-human games reaching $0.821$ Area Under the Curve (AUC). In the evaluation experiments, the SDMCTS agents played head-to-head cheat games against 120 human subjects. The results suggest that the SDMCTS agent performs well against humans, reaching a win ratio of $88.97\%$. Furthermore, its performances improve as the predictive model's accuracy increases. 

To conclude, the main novelties of this paper are: First, we present SDMCTS, an algorithm for combining prediction of human decisions with ISMCTS. Second, we introduce the Cheat Game as a test-bed for imperfect information games with partially observable actions. In addition, we present a highly efficient SDMCTS agent that performs well against humans and can run on a standard PC. Furthermore, we present a behavioral based prediction model of human decisions in the Cheat Game with $0.821$ AUC accuracy. Lastly, experimental results demonstrate the skillful performance of the SDMCTS approach.

%The Smooth-UCT was chosen as the online search algorithm. 

%Experiments with 80 subjects playing head-to-head games was conducted. The participants were divided into four groups. Each group played against a different variation of the algorithm. The results suggest that modelling partially observable action into Smooth-UCT search perform well in two-players games against humans. Moreover, the combination of the predictive model with the Smooth-UCT adaptation can further improve game performance.

%To conclude, the main novelties of this paper are: First, we present an algorithm for combining human prediction with MCTS algorithms. Second, we introduce the Cheat Game as a test-bed for imperfect information games with partially observable actions.
%In addition, we present a highly efficient Smooth-UCT agent that performs well against humans and can run on a standard PC. Furthermore, we present a behavioral based prediction model of human decisions in the Cheat Game with good accuracy (~0.82 AUC). Lastly, experimental results demonstrate the skillful performance of the \textit{Cheat Game} agent (MGA) and the combined predictive model with Smooth-UCT (PMGA).

\section{Background and Related Work}

%In the following sub-sections a brief overview of extensive-form games, MCTS, Smooth-UCT and predicting human decision is presented.

\subsection{Extensive-Form Games}

In this section a brief overview of the game tree representation and its formal notation are presented. An extensive-form game is a tree-based representation of a sequential interaction of multiple fully rational players. 
The set of players is denoted by ${\mathcal{N}} = \{ 1, ..., n \}$. In addition, a special player commonly called Chance is denoted by $c$ which has a probability distribution over its actions.
The set of all possible game states is denoted by ${\mathcal{S}}$. Each $s \in \mathcal{S}$
corresponds to a tree node in the game tree representation. A reward function 
$R : {\mathcal{S}} \to \Re^n$ maps the terminal states to the corresponding payoff vector where $R(s)_i$ is player $i$'s payoff. Each player's goal is to maximize his payoff in the game. 
The function $P : {\mathcal{S}} \to {\mathcal{N}} \cup \{c\}$ returns the player that is allowed to act in that state. The set of all available actions to player $P(s)$ is denoted by ${\mathcal{A}}(s)$. Each action $a \in {\mathcal{A}}(s)$ corresponds to an edge from the current state $s$ to a successor state. As mentioned above, in imperfect information games the game state is not fully observable by the players. The Information State ${u^i \in \mathcal{U}}^i$ is the observable portion of a game state $s$ for player $i$.
Thereby, the \textit{Information Function}  is defined as ${\mathcal{I}}^i : S \to {\mathcal{U}}^i$. It is important to note that
in cases where multiple game states are indistinguishable by the player, the game states will be mapped to the same information state. Therefore, the set of indistinguishable game states that corresponds to an information state $u^i$ is denoted by ${\mathcal{I}}^{-1}(u^i) = \{ s\: \rvert \: {\mathcal{I}}(s) = u^i \}$.

%A game is said to be \textit{perfect recall} if the information state $u_i^k$ implies knowledge of the sequence of their information states and actions, $u^i_1, a^i_1, u^i_2, a^i_2, ..., u^i_k$, that led to $u^i_k$. Although imperfect information games with partially observable actions does not satisfy this requirement, the method presented in this paper regards hidden actions as public information by revealing the action to all players. Therefore, the method's representation qualifies as perfect recall (a more detailed explanation is presented later in section \ref{sec:PredMCTS}.

The behavioural strategy of player $i$, denoted by $$\pi^i(u) \in \Delta (A(u)), \forall u \in U^i$$, is a probability distribution over the set of available action ${\mathcal{A}}(u^i)$. ${\Pi}^i$ denotes the set set of all behavioural strategies of player $i$. Furthermore, a strategy profile $\pi = (\pi^1, ... , \pi^n)$ is the set of all players' strategies. $\pi^{-i}$ denotes all strategies in $\pi$ except $\pi^i$. The expected payoff of player $i$ where all players follow strategy profile $\pi$ is denoted by $R_i(\pi)$. A strategy $\pi^i$ is said to be \textbf{best response} if for a given fixed strategy profile $\pi^{-i}$, $R_i(\pi^{-i} \cup \pi^i)$ is optimal. \textit{A Nash equilibrium} is a strategy profile $\pi$ such that for each player $i \in \mathcal{N}$, the strategy $\pi^i \in \pi$ is best response to $\pi^{-i}$.

\subsection{Monte Carlo Tree Search}\label{sec:MCTS}

MCTS \cite{coulom2006efficient} is a simulation-based search algorithm for finding optimal strategies. The algorithm was designed for a high-dimensional search space. Each simulated path through the game tree is selected to optimize exploration in more auspicious regions of the search space. One of the more interesting properties of MCTS is that it guarantees to converge to optimal strategies for perfect-information two-player zero-sum games \cite{kocsis2006bandit}. MCTS is comprised of four main components. The first component is a method for advancing the game and is denoted by $s_{t+1} \leftarrow {\mathcal{G}}(s_t, a_t)$. The second is a mechanism for updating statistics for each visited node (i.e. game state). The third component is an action selection method that is based on statistics stored for visited tree nodes. The last component is the rollout policy. The rollout policy is used to select actions when the simulation encounters game states that are out of the scope of the game tree. That is, a default strategy for selecting actions when there is no tree node that corresponds to the game state in the search tree. A detailed explanation of the Cheat Game MCTS representation is presented later in the section \textit{The Cheat Game Agent}.

For a given computational budget, MCTS performs multiple simulations.
Each simulation starts at the initial game state and traverses the game tree using the above action selection mechanisms. When a simulation reaches a terminal state, the leaf node's payoff is propagated back to all visited nodes and the algorithm updates their statistics. For each game state $s_t$ the algorithm stores the following statistics:
$N(s)$ denotes the number of times node $s$ has been visited. $N(s, a)$ denotes the number of times action $a$ was chosen when visiting node $s$. Lastly, $Q(s, a)$ denotes the aggregated discounted payoff for performing action $a$ at state $s$.

\subsection{Determinization}

Many MCTS variations were extended to imperfect information games \cite{browne2012survey}.
One of the more popular approaches performs the search on a determinized instance of the game \cite{ginsberg2001gib,sheppard2002world,sturtevant2008analysis,bjornsson2009cadiaplayer}. An imperfect information game can be converted into a perfect information game (i.e. a deterministic  game) by making the game states fully observable by all players and fixing the outcomes of stochastic events. A determinized MCTS algorithm performs simulations on the deterministic game with perfect information. In order to obtain a strategy for the full imperfect information game, the algorithm samples multiple determinized states that correspond to the current information state, and for each action uniformly averages its discounted reward (i.e. $Q(s, a)$). However, determinization techniques are susceptible to several problems which prevent its convergence to Nash-Equilibrium \cite{long2010understanding}. One of the more commonly reported problems is \textit{strategy
fusion} \cite{frank1998finding}. Strategy fusion occurs when performing simulations of a determinized perfect information game which relies on the assumption that a player can perform distinct actions on multiple game states. However, due to imperfect information, these game states may be indistinguishable by the player and the error will be propagated and distort the discounted reward estimations in earlier rounds. Information set MCTS \cite{cowling2012information} is a MCTS variation that was developed to address this problem. 

\subsection{Information Set MCTS}\label{sec:UCT}

Information set MCTS (ISMCTS) \cite{cowling2012information} is a MCTS variation that performs simulations directly on trees of information sets. Each node in the tree represents an information set from the point of view of the player and the statistics stored accordingly. However, when performing an online search, the algorithm is vulnerable to \textit{Non-Locality}. Non-locality occurs due to optimal payoffs not being recursively defined over subgames as in perfect information games. 
As a result, guarantees normally provided by search algorithms built on subgame decomposition no longer hold. Online Outcome Sampling (OOS) \cite{lisy2015online} is a variant of Monte Carlo counterfactual regret minimization (MCCFR) \cite{lanctot2009monte}. It is the first MCTS approach that addresses these problems by performing each simulation from the root state. While OOS guarantees convergence to Nash equilibrium \textbf{over time} in all two-player zero-sum games, other MCTS variations may yield better performance in certain situations. Smooth-UCT \cite{heinrich2015smooth} is a variant of the established \textit{Upper Confidence Bounds} Applied to Trees (UCT) algorithm \cite{kocsis2006bandit}. Similar to fictitious play, Smooth-UCT's action selection mechanism is designed to mix a player's average policy during self-play. Smooth-UCT requires the same information as UCT (i.e. node statistics). However, Smooth-UCT uses the average strategy by utilizing the $N(u,a)$ value. More specifically, Smooth-UCT chooses the average strategy with probability $\eta_k$ and the standard UCT action with probability $1 - \eta_k$,  where $\eta_k$ is an iteration sequence with $\displaystyle\lim_{k\to\infty}{{\eta}_k}=\gamma>0$. Smooth-UCT converges much faster than OOS to a sub-optimal strategy and was only outperformed by OOS after a significant number of simulated episodes \cite{heinrich2015smooth}. This suggests that for games with time constraints, Smooth-UCT outperforms better than OOS . For this reason, Smooth-UCT was selected as the online MCTS algorithm in the experiments. It is important to note that the present technique is independent of the online MCTS variation and can be extended to any Online Information set MCTS variation. The section \textit{Smooth-UCT Calibration} explores in detail the method for calibrating the Smooth-UCT parameters that were used for the experiments.

\subsection{Related Work on Predicting Human Decisions}

While many games have an optimal strategy for playing against fully rational opponents, empirical studies suggest that people rarely converge to the sub-game perfect equilibrium \cite{erev1998predicting,gal2011adaptive}. 
However, in many settings, it is still possible to develop a general human decision-making model for predicting human decisions using data collected from other people \cite{rosenfeld2016providing,rosenfeld2015learning,rosenfeld2012combining}. On the other hand, generalizing computational agents' decision-making without prior knowledge of their computational model is significantly harder \cite{demiris2007prediction}. Therefore, an integrative approach that incorporates prediction of a human opponent's decision-making may yield better game performance when interacting with an opponent for the first time, especially when dealing with uncertainty and deceptiveness. By applying prediction of human decision-making the agent adapts its actions to the human player and can better plan its future actions \cite{davidson2000improved,gal2004learning,markovitch2005learning,rosenfeld2012combining}. Facial expressions were used for predicting people's strategic decisions in the \textit{Centipede-Game} \cite{peled2013predicting}. Key facial points were extracted from video snippets of the players' faces and were used to train a classifier to predict participants' decisions. \cite{rosenfeld2016providing} presented the Predictive and Relevance based Heuristic agent (PRH), which can assist people in argumentative discussions. The agent utilized a predictive model with $76\%$ accuracy of people's top three arguments in conjunction with a heuristic model. \cite{he2016opponent} introduced neural-based models that learn
a policy and the behavior of the opponents. An encoded observation of the opponent's action is integrated into a deep Q-Network (DQN). The presented methods outperform DQN in both a simulated soccer game and a trivia game.

\section{Semi-Determinized MCTS}\label{sec:PredMCTS}

We introduce the Semi-Determinized MCTS (SDMCTS), a variant of the Information Set MCTS algorithm (ISMCTS).  
SDMCTS utilizes a predictive model of the unobservable portion of the human opponent's actions. First, SDMCTS searches for an optimal strategy as a response to the opponent's possible strategies in an instance of the game where hidden actions are revealed to all players. Second, SDMCTS uses the predictive  model  to  estimate  the  expected  reward  for  each  response  action.
%SDMCTS mixes between determination and Information Set MCTS. 
More specifically, for a given computational budget, SDMCTS performs Monte Carlo simulations directly on the information set of an instance of the game where 
\textbf{only} the unobservable portion of the opponent's actions are determined.
That is, during the simulations, SDMCTS considers all actions to be public information and are revealed to all players while the remaining private information is kept hidden.
Formally, let $u^i_k$ be the information state for player $i$ at round $k$ of the game. We denote by $a^o_{k-1}$ the action performed by an opponent in the previous round $k-1$ which led to the information state $u^i_{k}$. Note that $a^o_{k-1}$ is not fully observable by player $i$. Therefore, we define an alternative semi-determinized information state $\tilde{u}^i_k(a^o_{k-1})$ where the opponent's previous action is determined to be $a^o_{k-1}$. 
%While standard ISMCTS game trees contain the payoff estimations $Q(u^i,a)$ for performing response action $a$ at the information state $u^i$, t
During the search phase, SDMCTS performs simulations on the semi-determinized information states, resulting in estimates of $Q(\tilde{u}^i_k(a^o_{k-1}), a)$ for performing response action $a$ at the information state $u^i_k$ where the opponent's previous action is determined to be $a^o_{k-1}$ \footnote{The $k$ and $k-1$ notation are omitted in places where they can be deduced from the context}.

Once the search phase has been completed, the predictive model is used for calculating the expected payoff for each of the current player's response actions. The predictive model provides a probability distribution over the opponent's possible previous actions. Formally, for a given information state $u^i$, the predictive model estimates ${\mathcal{P}}(a^o \given u^i)$, the probability that the opponent has performed action $a^o$ in the previous round. The expected payoff ${\mathbb E}_u[u^i, a]$ for performing action $a$ at information state $u^i$ is calculated by:

$${\mathbb E}_u[u^i, a] = \sum_{a^o} \Big( {\mathcal{P}}(a^o \given u^i) \cdot Q\big(\tilde{u}^i(a^o), a\big) \Big)$$

where $Q(\tilde{u}^i(a^o), a)$ is the estimated payoff for performing response action $a$ at the semi-determinized information state $\tilde{u}^i(a^o)$. Lastly, the algorithm chooses the response action that maximizes the expected payoff. Algorithm \ref{alg} presents a pseudo code of the proposed method. In the algorithm, we denote by ${\mathcal{Q}}_{tree}$ the estimated payoff function that results from the SDMCTS simulations on the semi-determinized instance of the game.

\begin{algorithm}
\caption{Semi-Determinized ISMCTS}\label{alg}
\begin{algorithmic}[1]
\Function{SEARCH}{$u^i$} 
    \While{\textit{Within Computational Budget}}
        \For{\textbf{each} \textit{opponent's action} \textbf{$a^o$}}
        
            \State $s_0 \sim {\mathcal{I}}^{-1}(\tilde{u}^i(a^o))$
            \State{SIMULATE$(s_0)$}
         \EndFor
    \State \textbf{end for}
    \EndWhile
    \State \textbf{end while}
    \State \Return ${\mathcal{Q}}_{tree}$
\EndFunction{\textbf{end function}}
\Function{GET BEST RESPONSE}{$u^i$} 
    \State ${\mathcal{Q}}_{tree} \gets SEARCH(u^i)$
    \For{\textbf{each} $a \in {\mathcal{A}}(u^i)$}
        \State $${\mathbb E}_u[u^i,a] = \sum_{a^o} \Big( {\mathcal{P}}(a^o \given u^i) \cdot Q\big(\tilde{u}^i(a^o), a\big) \Big)$$
     \EndFor
   \State \Return $ \argmax_{a \in {\mathcal{A}}(u^i)}  {\mathbb E}_u[u^i,a] $
\EndFunction{\textbf{end function}}
\end{algorithmic}
\end{algorithm}

During the SEARCH, SDMCTS performs multiple ISMCTS simulations on the semi-determinized instance of the game. Algorithm \ref{algISMCTS} describes the standard ISMCTS simulation function. As described in section \ref{sec:MCTS}, ISMCTS is comprised of the game simulator $\mathcal{G}$ and the reward function $\mathcal{R}$.
In addition, \textit{OUT-OF-TREE} keeps track of players who have left the scope of their search tree in the current iteration (episode), i.e. the simulation has reached Information Set nodes that were not explored before. When the simulation encounters a new Information Set, one that was not visited in a/the previous episode, the \textit{OUT-OF-TREE} indicator is set to \textit{true} and a new tree node is created for the newly encountered Information Set. This operation is performed by the EXPANDTREE function.
The action selection and node updating functions (i.e. SELECT and UPDATE) determine the specific ISMCTS variant. The action selection function samples the tree policy and chooses an action for the specified information set. The update function is responsible for updating the tree nodes, i.e. updating the information set statistics. The specific implementation of these functions is derived from the choice of the ISMCTS variant. For the Cheat Game agent implementation which was used in the experiments, the Smooth-UCT was used. The implementation of the UPDATE and SELECT functions as well as the exact parameters' values that were used for the experiments are presented later in section \ref{sec:cal}.

\begin{algorithm}
\caption{ISMCTS Simulation}
\label{algISMCTS}
\begin{algorithmic}[1]
\Function{ROLLOUT}{$s$} 
    \State $a \sim \pi_r(s)$
    \State $s' \sim \mathcal{G}(s,a)$
    \State \Return SIMULATE($s'$)
\EndFunction{\textbf{end function}}
\Function{SIMULATE}{$s$} 
    \If{ISTERMINAL($s$)}
        \State \Return $r \sim \mathcal{R}(s)$ 
    \EndIf{\textbf{end if}}
    \State{$i=PLAYER(s)$}
    \If{OUT-OF-TREE($i$)}
        \State{\Return ROLLOUT($s$)}
    \EndIf{\textbf{end if}}
    \State{$u^i = \mathcal{I}^i(s)$}
    \If{$u^i \notin T^i$}
        \State{EXPANDTREE($T^i, u^i$)}
        \State{$a \sim \pi_r(s)$}
        \State{OUT-OF-TREE($i$) $\gets$ \textbf{true}}
    \Else
        \State{$a=SELECT(u^i)$}
    \EndIf{\textbf{end if}}
    \State{$s' \sim \mathcal{G}(s,a)$}
    \State{$r \gets SIMULATE(s')$}
    \State{UPDATE($u^i, a, r$)}
    \State{\Return r}
\EndFunction{\textbf{end function}}
\end{algorithmic}
\end{algorithm}

\section{The Cheat Game}

The \textit{Cheat Game} is an Imperfect Information Game with partially unobservable actions. We use it for demonstration of our \textit{SDMCTS}. In addition, we believe it is an excellent game for studying Imperfect Information scenarios and it is fun for people to play. We slightly adapted the Cheat Game to fit playing online. In our version of the Cheat Game, eight cards are dealt to each player at the onset of the game. The first player is chosen randomly. Play proceeds in the order of the deal. The objective of the game is to be the first player to get rid of all of his cards. A turn consists of a player placing a specific number (between one and four) of face-down cards in the middle of the table and making a claim as to what those cards' rank is. However, a player is permitted to deceive his opponent and lie about the cards' rank; we call this claim a false claim. The first claim of the game is chosen as the top card of the deck; subsequent calls must be exactly one rank higher or one rank lower, with kings being followed by aces. Lastly, if a player wishes to avoid making a claim, he may \textit{Take a Card} from the deck.
Once a player has made a claim, the opponent can challenge it by performing the Call-Cheat action. If a claim is challenged, the entire stack of cards that were placed onto the table are revealed and the claim is examined. If the challenge was correct, the player who made the false claim must take the entire stack of cards. However, if the challenger was wrong, he must take the stack. 
%Play continues in normal rotation as the next player starts a new pile. 
The first player to empty their hand is the winner. To further decrease the size of the search space, if the game does not conclude after 100 rounds, the games ends and the winner is the player who holds the lowest number of cards.% in her hand.

%Similar to Poker, 
The Cheat Game is an imperfect information game where a player cannot see his opponents' cards. In addition, as described above, the actual rank of a claim is not revealed to the other participants. In other words, the actions of a player are hidden and often deceptive. This property of the game adds a level of uncertainty and increases the complexity of finding a suitable strategy. 
%To address this form of an imperfect information game, the \textit{Predictive MCTS Cheat Game Agent} (PMCA) is proposed.

\subsection{The Cheat Game Agent}\label{sec:CheatMCTS}

The \textit{Predictive MCTS Cheat Game Agent} (PMCA) is an instantiation of the proposed method presented in the Section \textit{Semi-Determinized MCTS}. More specifically, the PMCA combines a variation of the MCTS algorithm with a predictive model of human decisions. As mentioned above, the chosen MCTS adaptation is the well-established Smooth-UCT \cite{heinrich2015smooth}. The predictive model was developed based on a human behavioral data-set. Prior to developing the agent, a preliminary experiment was conducted. 60 participants were asked to play the Cheat Game in a two-player repeated human-vs-human experiment.
%In addition, the agent was developed in collaboration with a highly experience \textit{Cheat Game} player. 

\subsection{The Cheat Game - Information State}

As defined above, an information state $u^i$ is the visible portion of the game state $s$ to player $i$. In the Cheat Game, the player is granted access to the following attributes of the game state. Clearly, the player can view his own cards and his own actions. In addition, the player can view the number of cards in the opponent's hand, the facedown table cards, and the remaining cards in the shuffled deck. Once a \textit{Call-Cheat} action is performed, the last claim that was made is examined. During the examination, all of the cards that were placed facedown on the table are revealed to all of the players. Therefore, a player can track cards that each player has collected from the table. 
%Furthermore, by examining the revealed cards with correlation to the opponent's action history, the player can deduce which of the opponent's past claims were false.
%Therefore, an information state $u^i_t$ at round $t$ can be determined from the initial information state $u^i_0$ and the players' action history $(a_1, a_2, ..., a_t)$ that led to $u^i_t$.

\subsection{The Cheat Game - Information State Abstraction}

An information state abstraction is a technique for significantly lowering the size of the state space. More specifically, an abstraction aggregates similar information states, resulting in an alternative information space
with a considerably smaller size \cite{johanson2016robust,gilpin2009algorithms}. Formally, an abstraction
$F = \{f_A^i : {\mathcal{U}}^i \to \tilde{\mathcal{U}}^i \vert i \in \mathcal{N}\}$ is a set of functions that maps the information state space ${\mathcal{U}}^i$ onto an information state space $\tilde{\mathcal{U}}^i$, where $\vert {\mathcal{U}}^i \vert \gg \vert \tilde{\mathcal{U}}^i \vert$.
While an abstraction is extremely important for reducing the search space, it is not without flaws. Aggregating similar information states prevents the players from distinguishing between the aggregated states and thus they may choose a sub-optimal action. Therefore, it is important to choose a suitable abstraction that both reduces state space and partially preserves its strategic structure.

The abstraction calibration was planned carefully to balance between state space size and preserving the strategic structure of the aggregated states. There is an inherent trade-off: 
as the search space decreases, the probability of aggregating information states with different strategies increases.  
Therefore, crucial attributes from the full information state $u^i$ were selected based on their importance when considering a strategy.  Multiple sub-sets of attributes were hand-picked by an experienced player and were tested extensively against human players. The sub-set of attributes that performed best was selected for the \textit{Cheat Game} information state abstraction. It is important to note that, as discussed above, during the search the proposed method regards hidden actions as public information. That is, the search is performed on the alternative information state where claim actions are determinized, i.e. claims are considered as public information and are revealed to all players. Therefore, the alternative information state contains public information about the nature of the claim, i.e. whether the previous claim is a \textit{true} claim or a \textit{false} claim. The information state abstraction is comprised of the following attributes:

\begin{center}
\begin{tabular} {|m{5mm}|| m{45mm} | m{18mm} ||}
 \hline
\textbf{\#} & \textbf{Description} & \textbf{Values} \\
 \hline
 \hline
 1 & The opponent's previous action. & \textit{TrueClaim}, \textit{FalseClaim}, \textit{TakeCard}, \textit{CallCheat}  \\
\hline
 2\& 3 & Can the player make a one card \textbf{higher/lower} true claim? & \textit{True} / \textit{False}\\
\hline
 4 & The player's card count. & $\psi\{1, ..., 52\}$\\
\hline
 5 & The opponent's card count. & $\psi\{1, ..., 52\}$\\
\hline
 6 & Placed on table card count. & $\psi\{1, ..., 52\}$\\
\hline
 7 & The round index. & $\{1, ..., 100\}$\\
\hline
\end{tabular}
Where $\psi(x)= \begin{cases}
    x,& x\leq 4\\
    5,& 5 \leq x \leq 8\\
    6,              & \text{otherwise}
\end{cases}$
\end{center}

The remaining attributes require some additional explanation. For the duration of the game, the agents keep track of which cards were collected by the opponent when Call-Cheat was performed. In this way, the agent can estimate which cards are currently held by the opponent. In addition, the agents keep track of the opponent's claims. Some of the abstraction's attributes are derived from these estimations.

\begin{center}
\begin{tabular} {|m{5mm}|| m{45mm} | m{18mm} ||}

 \hline
 8 \& 9 & The estimated number of cards the opponent has that are one rank \textbf{higher/lower} from the last valid claim. & $\varphi\{1, ..., 4\}$
 \\
\hline
10 \& 11 & A value indicating whether the current player has made the same \textbf{higher/lower} claim since the last \textit{Call-Cheat} move. & \textit{True} / \textit{False} \\
\hline
12 \& 13 & A value indicating whether the opponent has made the same \textbf{higher/lower} claim since the last \textit{Call-Cheat} move. & \textit{True} / \textit{False} \\
\hline
14 \& 15 & A value indicating whether the current player was caught cheating on a claim with one rank \textbf{higher/lower} than the last claimed rank. & \textit{True} / \textit{False} \\
\hline
16 & Did the opponent catch the player cheating on any rank? & \textit{True} / \textit{False} \\
\hline
\end{tabular}
Where $\varphi(x)=\begin{cases}
    x,& \text{if } x\leq 2\\
    3,& \text{otherwise}
\end{cases}$
\end{center}

\subsubsection{Satisfying Time Restrictions}

\begin{figure}[h]
\begin{center}
\includegraphics[width=0.9475\columnwidth]{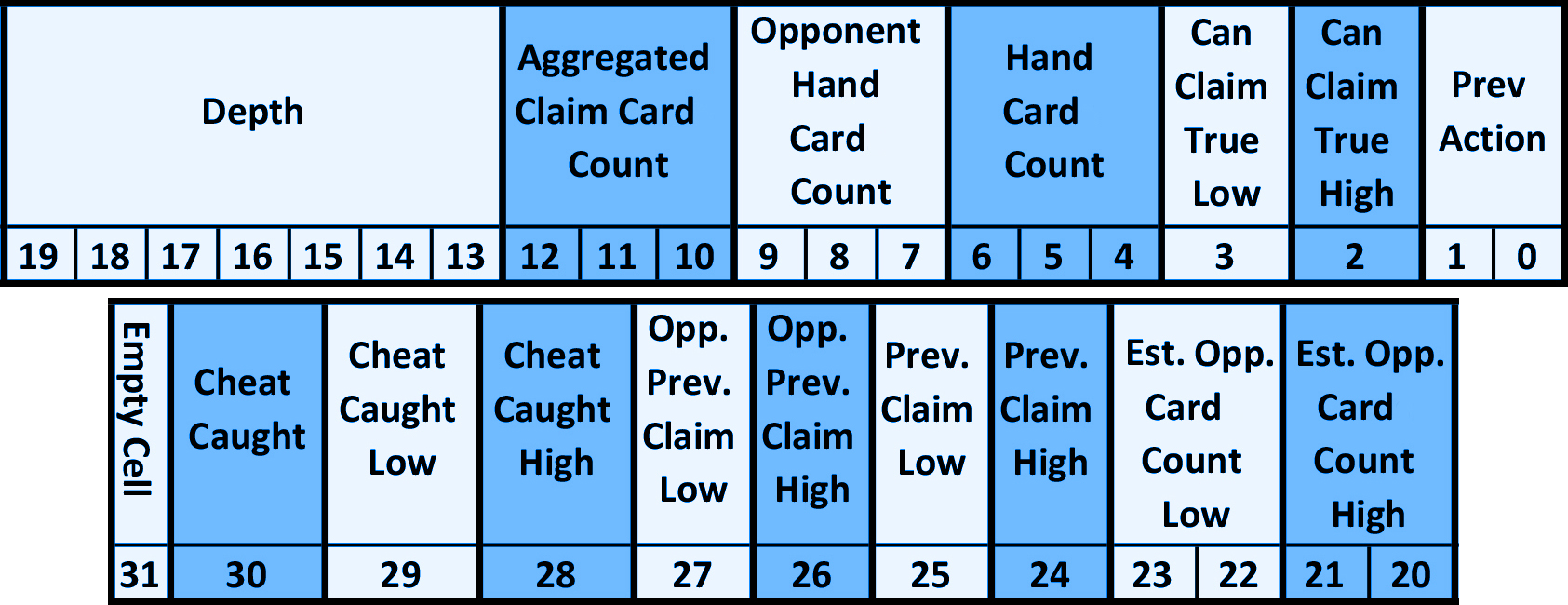}
\end{center}
\caption{Optimization - 32-bit Encoding}
\label{fig:Opt}
\end{figure}

The time duration for producing an action for the agents was determined from the average time it took the participants to respond when they played head-to-head human vs human. This value was determined to be 25 seconds.
In order to create an agent that can produce a response in reasonable time, the MCTS tree nodes were optimized. While a naive representation of the described attributes takes 15 bytes in size, we were able to encode all of the attributes into a 32-bit uint structure without data loss (see Figure \ref{fig:Opt}). This encoding was done for two reasons. First, the encoding ensures a low memory footprint. Second, the use of a 32-bit uint is extremely suitable for hash-mapping structures. For a large enough heap, it can significantly reduce the conflict when hash retrieval is called and thereby improves CPU usage. In particular, for the same amount of time, we were able to achieve a 12 fold improvement in the number of simulations that can be performed.

\subsection{The Cheat Game - Action Abstraction}

For a given game state $s$, a player may perform the following actions.
If the opponent made a claim during the previous state, the player may choose to accuse the opponent of ``cheating'', i.e. perform the \textit{Call-Cheat} action.
Alternatively, a player may make his own claim. As mentioned above, a claim can be either true or false. The player may place up to four cards facedown on the table and make a claim as to their rank. Lastly, if a player wishes to avoid making a claim, he may \textit{Take a Card} from the deck. Therefore, for a given game state $s$, the number of actions a player may perform is:
$$\vert{{\mathcal{A}}(s)}\vert = 2 + \sum_{k=1}^{4} {h \choose k}$$
where $h$ is the number of cards in the player's hand and $k$ is the number of cards he wishes to declare. The additional two actions are for the \textit{Call-Cheat} and \textit{Take a Card} moves. Consequently, for an average hand of eight cards, the  branching factor will result in $164$ unique actions. Combined with the 100 turn limit, the estimated game tree may contain up to $10^{222}\approx164^{100}$ nodes. For this reason, an action abstraction is introduced. A claim action is represented by a tuple $(\gamma, \delta, k)$, where
$\gamma \in \{higher, lower\}$ denotes the direction of the claim, $\delta \in \{True, False\}$ is a value indicating whether the claim is true, and $k$ is the number of cards that was declared. In return, the branching factor is significantly reduced to a more computationally manageable size, that is: $\vert{\mathcal{A}}(s)\vert = 2+\vert \{(\gamma, \delta, k)\}\vert = 18$.

\subsection{The Cheat Game - Rollout Policy}

\begin{figure}[h]
\begin{center}
\includegraphics[width=\columnwidth]{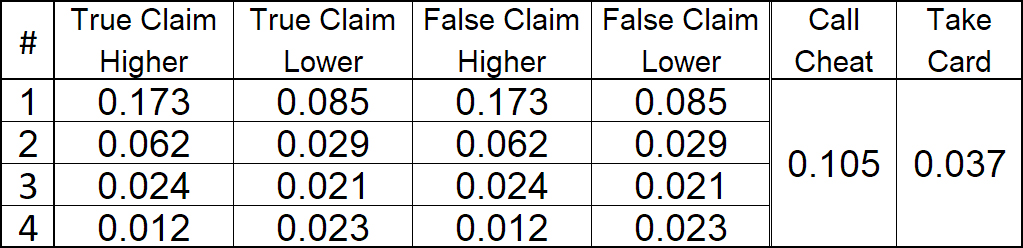}
\end{center}
\caption{Rollout Policy PDF over the action abstraction}
\label{fig:Roll}
\end{figure}

A range of unique heuristic based rollout policies were tested against experienced human players. The rollout policy that performed best against these experienced players was selected for the MCTS simulations. The \textit{rollout policy}, denoted by $\pi_r$, was designed to mimic the average strategic behavior of human players. To that end, the human-vs-human data-set was used to extract a probability distribution function (PDF) over the available actions. The probability distribution function was combined with a set of heuristic rules which were designed to handle special cases. The probabilities over the set of action abstraction are presented in Figure \ref{fig:Roll}. In cases when one or more of the actions were inapplicable, the inapplicable actions' probability values were distributed uniformly among the remaining actions. For example, if the previous action was \textbf{not} a claim then the player was not allowed to perform the \textit{Call-Cheat} action. As a result, the remaining valid actions' probabilities would have increased by $0.006 = 0.105 / 17$ each. Formally, let $u^i$ be an information state and let $P_{r}(a)$ be the fixed rollout probability for choosing action $a \in {\mathcal{A}}(u^i)$, the normalized rollout probability for choosing $a$ in information state $u^i$ is calculated by: 
$$\pi_r(a \given u^i) \sim \frac{{\mathcal{P}}_r(a)}{\sum_{a_i\in {\mathcal{A}}(u^i)} {\mathcal{P}}_r(a_i)} $$

The special case heuristic was developed based on the expertise of the human player. It is important to note that the heuristic rules take precedence over the PDF. Therefore, if the current game state satisfies any of the special case's conditions, the heuristic action is performed and the PDF action is ignored. The set of rules was designed to reduce the chance that the rollout policy would perform a dominated strategy. If the opponent made a false claim, the rollout policy would choose the \textit{Call-Cheat} action for the following cases:
a) If the opponent has no more cards in his hand; b) If there are more than eight cards on the table; c) If the opponent was caught cheating more than eight times; d) If the opponent made the same claim since the last time Call-Cheat was performed; e) If the current player's cards contain a card with the same rank as the claim. f) If the opponent was caught cheating on the same rank in previous rounds.

%\begin{table}[htbp]
%\centering
%\begin{center}
%\begin{tabular}{|p{4mm}|p{9mm}|p{9mm}|p{8mm}|p{8mm}||p{8mm}|p{6mm}|}
%\hline
%\textbf{\#} & \textbf{True Claim Higher} & \textbf{False Claim Higher}& \textbf{True Claim Lower} & \textbf{False Claim %Lower} & \textbf{Call-Cheat} & \textbf{Take Card} \\ 
%\hline
%1 & 0.173 & 0.085 & 0.173 & 0.085 & \multirow{4}{*}{0.105} & \multirow{4}{*}{0.037}\\ 
%3 & 0.062 & 0.029 & 0.062 & 0.029 &&\\
%2 & 0.024 & 0.021 & 0.024 & 0.021 &&\\
%4 & 0.012 & 0.023 & 0.012 & 0.023 &&\\
%
%\hline
%\end{tabular}
%\end{center}
%\caption{rollout policy PDF over the action abstraction}
%\label{table:prob}
%\end{table}

\subsection{Smooth-UCT Calibration}\label{sec:cal}

The Smooth-UCT mixes the average strategy with the \textit{Upper-Confidence Bound (UCB)} action selection mechanism. More specifically, Smooth-UCT chooses with probability $\eta_k$ the UCB action:
$$a = argmax_{a} Q(u^i, a) + c \cdot \sqrt{\frac{log N(u^i)}{N(u^i, a)}}$$ where $c$ is the balancing factor between exploration and exploitation.
Alternatively, Smooth-UCT chooses with a probability of $1 - \eta_k$ the average strategy: 
$$a \sim p\text{, where }\forall a \in {\mathcal{A}}(u^i)\text{ : }p(a) \gets \frac{N(u^i,a)}{N(u^i)}$$.

The $\eta_k$ is an iteration k-adapted sequence defined as:
$$\eta_k =  max\bigg(\gamma, \eta \cdot \Big(1 + d \cdot \sqrt{N_k} \Big)^{-1}\bigg)$$
where $\gamma, \eta$ and $d$ are constants and $N_k$ is the total visits to the corresponding $u^i$ node.
(The interested reader is referred to \cite{heinrich2015smooth} for an in-depth explanation of the parameters and their importance).

The $\eta$ and $\gamma$ parameters were manually calibrated and set to $\eta =0.9, \gamma = 0.1$. The $d$, $c$ and payoff \textit{discount-factor} parameters were calibrated using a grid-search self-play tournament. Different combinations of the parameters were competing in a head-to-head tournament. Each game was comprised of 100 matches, the set of parameters who won the majority of matches continued through to the next level, while the loser was eliminated. The parameters' settings were $c = 17 + k$, $d = 0.001 + 0.005l$ and the \textit{discount-factor} $= 0.97 + 0.005m$ where $k,l,m \in \{1, ..., 10\}$.
The winning parameters set, $c = 0.0025, d = 0.0025$ and  \textit{Discount Factor} $= 0.995$, was used in the experiment.
 
Algorithm \ref{algISmooth-UCT} describes the action selection mechanism and update functions used in Smooth-UCT. It is important to note that Smooth-UCT uses the same UPDATE function as standard UCT. 

\begin{algorithm}
\caption{Smooth-UCT - Select \& Update}
\label{algISmooth-UCT}
\begin{algorithmic}[1]
\Function{SELECT}{$u^i$} 
    \State{$z \sim  U[0,1]$}
    \If{$z < \eta_k(u^i)$}
        \State{\Return $argmax_{a} Q(u^i, a) + c \cdot \sqrt{\frac{log N(u^i)}{N(u^i, a)}}$}
    \Else
        \State{$\forall a \in A(u^i):p(a) \gets \frac{N(u^i,a)}{N(u^i)}$}
        \State{\Return $a \sim p$}
    \EndIf{\textbf{end if}}
\EndFunction{\textbf{end function}}

\Function{UPDATE}{$u^i,a,r$}
    \State{$N(u^i) \gets N(u^i) + 1$}
    \State{$N(u^i,a) \gets N(u^i,a) + 1$}
    \State{$\mathcal{Q}(u^i,a) \gets \mathcal{Q}(u^i,a) + \frac{r-\mathcal{Q}(u^i,a)}{N(u^i),a)}$}
\EndFunction{\textbf{end function}}
\end{algorithmic}
\end{algorithm}

\subsection{Predictive Model}\label{sec:pred}

\begin{figure*}[!thb] \centering
		\begin{subfigure}[h]{0.23\textwidth}
			\frame{\includegraphics[width=\textwidth]{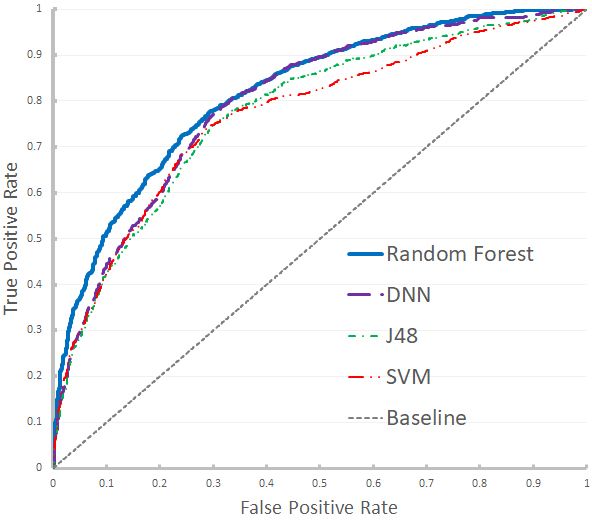}}
			\caption{ROC Curves}
			\label{fig:ROC}
		\end{subfigure}\hspace{0.01\textwidth}
		\begin{subfigure}[h]{0.23\textwidth}
			\frame{\includegraphics[width=\textwidth]{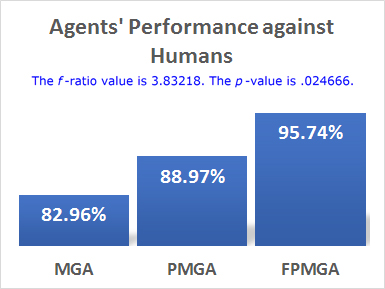}}
			\caption{Agents' matches win ratio}
			\label{fig:r1}
		\end{subfigure}\hspace{0.01\textwidth}
		\begin{subfigure}[h]{0.23\textwidth}
			\frame{\includegraphics[width=\textwidth]{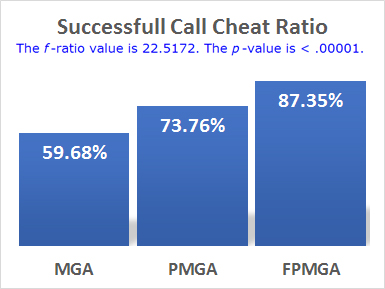}}
			\caption{Call Cheat Success Rate}
			\label{fig:r2}
		\end{subfigure}\hspace{0.01\textwidth}
		\begin{subfigure}[h]{0.23\textwidth}
			\frame{\includegraphics[width=\textwidth]{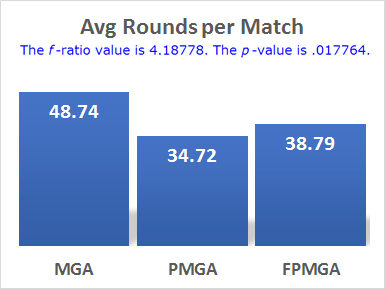}}
			\caption{Matches' Avg. Rounds}
			\label{fig:r3}
		\end{subfigure}
		\caption{Predictive Model \& Experimental results.}\label{fig:Experimental}
	\end{figure*}

The predictive model was devised to classify the opponent's action type, i.e. $true$ or $false$ claim. As mentioned in Section \textit{The Cheat Game Agent}, a data-set of human-vs-human play of 60 participants was collected in the preliminary experiment. The genders were distributed evenly with $51\%$ males and $49\%$ females with ages varying between 18-42. In total, the data-set contains 1,275 $true$ claim samples and 1,157 $false$ claim samples. The data-set was used to manually extract a collection of features. Customarily, different subsets of features were tested on a wide range of binary classifiers ranging from Decision Trees through Support Vector Machine (SVM) \cite{cortes1995support} to Deep Neural Network (DNN) \cite{schmidhuber2015deep}. The evaluation of the model was based on a “leave-one-sample-out” cross validation. For each sample for player $i$, round $t$ and match $m$, the classifier was allowed to train on samples from all players excluding player $i$ and his opponent, as well as samples of player $i$ and his opponent that precede match $m$ and round $t$. In order to avoid class imbalance, oversampling was used in cases where the number of $true$ and $false$ claim samples were significantly imbalanced. 
We were able to achieve a good prediction rate with an \textit{Area Under the Curve} (AUC) of $0.821$, $TPR = 75.3\%$, $TNR = 74.0\%$ and $G-Mean =  74.6\%$ using the \textit{Random Forest} algorithm. The Deep Neural Network (DNN) classifier received similar but slightly lower accuracy, however with no significant difference. Figure \ref{fig:ROC} presents the \textit{Receiver Operating Characteristic (ROC)} curves of the most accurate classifiers.
The final features collection contained 21 mostly statistical features.
The human player' behavioural changes were modeled using these predictive model's features. The features contains information that correspond to the number of rounds played. 
One of the more significant features is the response duration. Another feature is the number of times a player was caught cheating with respect to the number of times he made a false claim until the last call cheat (i.e. \#caught / \#cheat). One more interesting feature is the number of times the player took cards from the deck with respect to the number of rounds played. 

It is important to note that features\footnote{For the complete feature list visit: https://goo.gl/VLy7NH} were derived strictly from data that is contained within the Information State.

% \begin{figure}[h]
% \begin{center}
% \includegraphics[width=0.65\columnwidth]{ROC.jpg}
% \end{center}
% \caption{ROC curve for the human-vs-human data-set.}
% \label{fig:ROC}
% \end{figure}
\section{Experiments}\label{sec:experiment}

Experiment were conducted in order to evaluate the effectiveness of the proposed method. 120 participants from the US, aged 20-50 (46\% females and 54\% males), were recruited using Amazon Mechanical Turk.  
%The participants' aging from 20-50 (46\% females and 54\% males) were asked to participate in a two-player computer Cheat Game experiment. 
The participants were asked to play a two-player Cheat Game for at least three matches but for no more than five matches. Prior to the game, participants were presented with instructions followed by a quiz which ensured that the rules were explained properly. The participants received payment on a per match basis and an additional payment for every match they won.
%. In order to incentivize the participants, they were informed that an additional bonus payment would be granted for every match they win.

The objective of the experiment is to evaluate the effectiveness of combining MCTS with a predictive model.
%of human decisions. 
To that end, participants were randomly divided into three groups. Each group played against a different instantiation of the proposed method (see Section \textit{Semi-Determinized MCTS}).
As described above, the Smooth-UCT was chosen as the MCTS algorithm. 
In the first group, the participants
%The first group 
played against a \textit{Smooth-UCT Cheat Game Agent} (\textit{MGA}) 
without a predictive model. The agent gives equal probability to true and false claims, i.e. ${\mathcal{P}}(a^o = \text{true claim} \given u^i) = {\mathcal{P}}(a^o = \text{false claim} \given u^i) = 0.5$ where $u^i$ is an information state for player $i$ where the previous opponent's action, $a^o$, is a claim action. 
Naturally, in information states where the opponent's previous action is not a claim (i.e. TakeCard or CallCheat) these probabilities are set to zero.
The second group played against the \textit{Predictive Smooth-UCT Cheat Game Agent} (\textbf{\textit{PMGA}}) that incorporates the predictive model from the \textit{Predictive Model} section. As described above, the predictive model was trained on data collected from a preliminary experiment were the participants played human-vs-human cheat game.
It is important to note the preliminary experiment's subjects did not participate in the agents' evaluation experiments.
In order to further demonstrate the effectiveness of combining a predictive model with MCTS, participants in the third group played against the \textbf{\textit{FPMGA}} agent, which had an unfair advantage. The \textit{FPMGA} agent was allowed % to predict with $85\%$ accuracy if the claim was true or false by allowing it
to peek into his opponent's real claim.  Therefore, participants in the third group were compensated due to this unfair advantage.
While the FPMGA agent was able to predicate the opponent's claim with perfect precision (i.e. $100\%$ prediction rate), it was restricted to a prediction rate of $85\%$.
The $85\%$ prediction rate was chosen as a plausible prediction rate that can be achieved when predicting human decisions. 
A higher prediction rate (above $85\%$) is extremely difficult to obtain when interacting with people. 
This is in part because of the inherent noise in the human decision-making process. 
For example, the same person may choose a different strategy in the same exact game state. For the experiments, we hypothesize that the \textit{FPMGA} agent will outperform the \textit{PMGA} agent and that both predictive agents will outperform the non-predictive agent, \textit{MGA}.

\subsection{Experimental Results}

% \begin{figure}[h]

% \begin{center}
% \includegraphics[width=0.75\columnwidth]{Results.jpg}
% \end{center}
% \caption{Agents' matches win ratio}
% \label{fig:Results}
% \end{figure}
To analyze the results, one-way ANOVA was conducted to compare the three conditions ($p < .05$). The exact $p$\textit{-value} and $f$\textit{-ratio} can be observed in the figures. 
As can be seen in Figure \ref{fig:Experimental}, both the predictive agent (PMGA) and the pseudo predictive agent (FPMGA) performed significantly better than the non-predictive agent (MGA). Furthermore, FPMGA performed significantly better than PMGA. Figure \ref{fig:r1} presents the percentage of won matches by the agents (i.e. won/played). This can be explained by the successful call cheat ratio that can be viewed in Figure \ref{fig:r2}. The results suggest that by combining the predictive model with the Smooth-UCT, the agents were able to reduce the uncertainty that was derived from hidden action. Thereby, the predictive agents were able to choose better response actions to deceptive claims. Figure \ref{fig:r3} can further demonstrate the improvement of game performance. The average number of rounds it took the predictive agents to conclude a match is significantly lower than the non-predictive agent. Interestingly, the average number of rounds it took the FPMGA to conclude a match is higher than the PMGA, despite FPMGA having a more accurate prediction rate. We hypothesize that this is due to fact that the FPMGA performs the call cheat action more than the PMGA and MGA. This prolongs the number of rounds needed to reach a terminal state.
However, the MCTS's discount factor can be calibrated in order to incentivize the agents to conclude the games faster.

Another measure for play-dominance is the average difference between the cards held by the agent and the humans.
The PMGA's and FPMGA's average card difference ($-3.51$ and $-4.56$, respectively) was significantly lower than MGA's ($-2.89$), with an $f$\textit{-ratio} of $6.06$ and a $p$\textit{-value} of $.002$. The importance of the statistically significant results is enhanced when considering the low number of matches in the experiments.
%, i.e. 421. For comparison, the number of hands that are needed to reach significance in Computer Poker is over 80,000. 
%Interestedly, we were unable to find a statistically significant difference in people's strategic behavior with respect to the different groups.

In addition to the statistically significant results, we offer as a discussion the human participants' behavioral statistics. It seems that people played differently against the different agents. Unfortunately, statistical significance of such behaviour was not obtained. We hypothesize that this is because of the inherent noise in human behavior. 
For example, people who played against the predictive agents demonstrate a reduction in performing false claims. More specifically, $40.4\%$ and $42.4\%$ of all people's claims were false when played against FPMGA and PMGA, respectively, while people who played against MGA lied more with a $44.8\%$ false claim ratio.
This can be explained by the fact that people tend to lie less when there is a high chance of being caught. On the other hand, people who played against MGA took less cards from the deck, i.e. chose to Take-Card for $23.9\%$ of the moves, while people who played against the FPMGA and PMGA performed the Take-Card action in $19.6\%$ and $20.8\%$ of moves, respectively. This is unexpected, as one would expect people to take a card from the deck in order to avoid making a false claim. Another important observation is that people who played against MGA perform better in later matches, that is, people won $13\%$ of the first two matches and $28\%$ of the last two matches. However, people who played against PMGA and FPMGA performed in a similar fashion across all matches, contributing to the overall success of the prediction-based methods.

\section{Conclusion}

We have presented an algorithm for the development of agents that performs well against human in Imperfect Information Games with Partially Observable Actions. More specifically, we have introduced Semi-Determinized MCTS (SDMCTS), an ISMCTS algorithm that combines a predictive model of the opponent's actions and an information set MCTS variant. The method builds on existing ISMCTS adaptations and determinization techniques, and takes advantage of the benefits of both approaches. We have presented the \textit{Cheat Game} as an indicative example of the effectiveness of the presented techniques. In addition, a predictive model was conferred and produced good accuracy for predicting human strategic decision-making in the Cheat Game. We have presented the \textit{MGA}, \textit{PMGA} and \textit{FPMGA} agents which applied the SDMCTS algorithm to the Cheat Game. The SDMCTS agents combined the predictive model with the Smooth-UCT, a ISMCTS variation, to yield skillful performance when used in a head-to-head game with human opponents. The results of an extensive experiment with 120 participants were presented. The participants played repeated head-to-head games against the \textit{MGA}, \textit{PMGA} and \textit{FPMGA} agents. The results suggest that the combination of a predictive model with MCTS can be used to improve game performance against humans. Furthermore, the agents' performance improved as the predictive model's accuracy increases. 

For future work, we intend to combine an additional predictive model of the opponent's response strategy in the Cheat Game. More specifically, we will develop a predictive model that estimates the probability that the opponent will \textit{Call Cheat} at a claim. In addition, we intend to implement the SDMCTS algorithm for two-player poker and develop a predictive model that estimates the strength of the opponent's hand compared to ours. Lastly, we intend to implement our method on Phantom games.
\newpage

\bibliographystyle{ACM-Reference-Format}  % do not change this line!
\bibliography{sample-bibliography}  % put name of your .bib file here

\end{document}